\def\b{\begin{equation}}
\def\e{\end{equation}}
\def\l{\left}
\def\r{\right}
\documentclass[prl,aps,twocolumn, showpacs, showkeys, floats]{revtex4}

\begin{document}

\title
{An Astrophysical Peek into Einstein's Static Universe}

\author{Abhas Mitra}

\email {amitra@barc.gov.in}
\affiliation { Theoretical Astrophysics Section, Bhabha Atomic Research Centre, Mumbai, India}


\date{\today}






\date{\today}
\begin{abstract}
Einstein introduced the cosmological  constant ($\Lambda$) as a fundamental parameter residing on the Left Hand Side (LHS) of his field equations  for obtaining a finite density ($\rho$) for his model of  static universe ( Einstein's static universe :ESU).
For the sake of physical insight, we derive here the metric for  ESU directly from Einstein equation, i.e., by considering both Einstein tensor $G_{ik}$ and energy momentum tensor $T_{ik}$. We find that in order that the fluid pressure and acceleration are {\em uniform} and finite despite
the presence of a coordinate singularity, the effective density $\rho_e = \rho + \Lambda/8 \pi =0$, where $\Lambda$ is the cosmological constant. Under weak energy condition, this would imply
$\rho = \Lambda =0$ for ESU. This means that if one would need to invoke a source of ``repulsive gravity'' in some model, (i) the model must be non-static, (ii) the repulsive gravity must be due to a ``quintessence'' or a ``dark energy'' fluid with negative pressure and appear  on the right hand side (RHS) of the Einstein equation through $T_{ij}$ rather than through a fundamental constant($\Lambda$) residing on the LHS of the same equation, and (iii)  energy density  of  both normal matter and the ``dark energy fluid'' should be time dependent.
In fact, the repulsive gravity would be due to a time independent fundamental constant $\Lambda$, it would be extremely difficult to understand why the associated energy density should be approximately $10^{120}$ times lower than the value predicted by quantum gravity. On the other hand, for a dark energy fluid whose energy density is time dependent, it would be much easier to understand such an extremely low present energy density: the original initial value of the energy density of the fluid could be equal to the quantum gravity value while the present low value is due to decay with time.
\end{abstract}
\pacs{04.40.-b, 04.40.Nr}

\keywords{Self-gravitating systems, Space time with fluids, Fractals}
\maketitle

\section {Introuduction}
The ``cosmological principle'' demands that center of the universe lies everywhere and any fundamental observer must see the universe as isotropic.
This  means that, the space must be ``maximally symmetric''\cite{1}  which demands that at each moment the metric of the space is same at all points and all directions\cite{2}. By such considerations one can directly obtain not only the metric for Einstein's static universe (ESU)
but the non-static Friedmann- Robertson- Walker (FRW) metric as well\cite{1,2} without using Einstein's equation 
 \b
G_{ik} = 8 \pi~ T_{ik}
\e
at all ($G=c=1$). The Einstein tensor $G_{ik}$, on the left hand side (LHS) of this equation, comprises only geometric quantities, the structure of space time
$g_{ik}$. On the other hand, the right hand side (RHS) of the same equation contains much more tangible quantity, $T_{ik}$, the energy momentum tensor
of matter generating the space time structure $g_{ik}$ or $G_{ik}$. Thus any constraint imposed on $g_{ik}$ from symmetry or other considerations must take its toll on the  admissible  form of $T_{ik}$  as well. This means that constraints imposed on the LHS of Einstein equation must be reflected
on the equation of state (EOS) of the fluid generating $G_{ik}$. This latter physics aspect however may remain hidden in a purely symmetry driven mathematical derivation of cosmological metric which does not invoke Einstein equation at all. And in this paper, we would like to explore the
tacit constraints imposed  on $T_{ik}$ by the requirements of the cosmological principle, ``there is no preferred center'', or ``center lies everywhere''.
 All the results obtained during this process will be  also
ratified by physical considerations as well. 

In effect we shall be dealing with spherically symmetric static solutions of Einstein equation with the inclusion of a cosmological constant $\Lambda$,
a problem which was probably first dealt by Weyl\cite{3}. Since then several authors have  considered such solutions and one particular case
is due to Tolman\cite{4,5}. In recent times several authors have shown keen interest in this important problem\cite{6,7}. But, most comprehensive and elaborate
treatment of this topic was due to Bohmer\cite{8}. Since then Bohmer and his coworkers have tried to find even newer aspects of this problem which
was earlier thought to be almost a dead topic\cite{9,10}. However such detail works mostly focused attention on isolated objects or stars where
there (i) must be a pressure gradient, (ii) in general, a density gradient, (iii) a natural boundary $R=R_b$ where fluid pressure must vanish, $p(R_b)
 =0$ and (iv) where a discuntinuity in the fluid density $\rho$ might also occur. Further in such generalized studies, $\Lambda$ was not considered as
 a fixed fundamental constant contrary to what was in Einstein's mind when it was first introduced.
 
 In contrast, the present study,  $\Lambda \equiv \Lambda_{Einstein}$ is a basic constant and there is no question of modifying the value of $\Lambda$
 according to the needs of a general fluid solution. And of course, there must not be any natural ``boundary'' at all for the universe so that
 Copernican principle remains valid. It will be found that both the  metric for a star and ESU have a ``coordinate singularity at $R= \alpha$,
 where $R$ is the luminosity distance and $\alpha$ is an appropriate integration constant. This singularity gets propagated in the hydrostatic balance
 equation as well. And in order to ensure that isotropic pressure gradient $p'$ does not blow up at this singularity, relativistic stars are constructed
 so that $R_b > \alpha$. Accordingly one can almost forget this singularity while studying static stars. But for the ESU one has to live with this singularity. And in this paper, we would study the behavior of physical quantities like pressure and acceleration despite the presence of this singularity. 
\section{Formulation}
We start with the assumption of spherical symmetry and  consider a general form a static metric \cite{2}
\b
ds^2 = e^\nu dt^2  - e^\lambda dR^2 - R^2 d\Omega^2
\e
where $d\Omega^2 = d\theta^2 + \sin^2 \theta d\phi^2$ and $R$ is the area coordinate. In particular, in a static universe, the luminosity distance turns out to be exactly $R$\cite{3}
\b
d_L = R
\e
We also assume the cosmic fluid to be perfect with
\b
T_{ik} = (\rho +p) u_i u_k - p g_{ik}
\e
where, $\rho$ is the fluid density, $p$ is the isotropic pressure, and $u^i$ is fluid 4-velocity.  As is clear from Eq.(1), at the beginning, we do not consider any cosmological constant $\Lambda$; and the $_0^0$ component of the Einstein equation reads
\b
8 \pi T_0^0 = 8 \pi \rho = - e^{-\lambda} \l({1\over R^2} - {\lambda'\over R}\r) + {1\over R^2}
\e
This can be integrated to yield
\b
e^{-\lambda} = 1 - {\alpha(R) \over R}
\e
where
\b
\alpha(R) = \int_0^R 8 \pi \rho R^2 dR
\e
Here the condition $\alpha(0) =0$ has been used to ensure that $e^\lambda$ is regular at $R=0$. 
Thus whether, it is the interior solution of a star or the static universe, general form of the metric is
\b
ds^2 = e^\nu dt^2 - {dR^2\over 1 - \alpha/R} - R^2 d\Omega^2
\e
Note, as of now, we need not necessarily interpret  $\alpha(R)$ as related to
observed mass though  what we have done looks like finding the interior of a relativistic star by working out the Schwarzschild interior solution.
The important difference, however, is that while for a star there is a unique center and a boundary where the density may be discontinuous, 
 for the universe, center is everywhere and there is no boundary, no exterior solution, and no density discontinuity.
Further  even for a constant density star, there must be a pressure gradient because there is a unique center. But for the universe {\em pressure gradient must vanish everywhere} because the {\em the center lies everywhere} and there is no boundary discontinuity. In fact for an isotropic
and homogeneous universe, not only $p$ and $\rho$, but all physically meaningful quantities, all geometrical scalars must the same everywhere.
Such requirements may demand severe restriction on the admissible equation of state (EOS) of the cosmic fluid.  In contrast for a star, at best only $\rho$
can be uniform in which case there would be a discontinuity at its boundary $R=R_b$. 

Now we introduce the condition that the fluid must be homogeneous: $\rho = uniform$ and thus both for the universe as well as a constant density
star, one has
\b
\alpha = {8 \pi\over 3} \rho R^3
\e
and the metric assumes the form
\b
ds^2 = e^\nu dt^2 -   {dR^2\over 1 - (8 \pi  /3) \rho R^2 } + R^2 ~d\Omega^2 
\e
If we  introduce a parameter
\b
S = \sqrt{3\over 8 \pi \rho}
\e
Eq.(10) would acquire the form 
\b
ds^2 = e^\nu dt^2 -   {dR^2\over 1 - R^2/S^2} + R^2 ~d\Omega^2 
\e
Further, if we introduce a new coordinate
\b
r = R/S
\e
we can rewrite the above metric as
\b
ds^2 = e^\nu dt^2 -  S^2 \l( {dr^2\over 1 - r^2} + r^2 ~d\Omega^2 \r)
\e
Hence, the spatial section of both the universe and the interior of a constant density star is that of a 3-sphere, a space of constant
curvature, a fact noted by Weyl\cite{4}. This becomes clearer if we express
\b
r = R/S = \sin \chi
\e
to rewrite Eq.(14) as
\b
ds^2 = e^\nu dt^2 -  S^2 ( d\chi^2 + \sin^2 \chi ~d\Omega^2) 
\e
The coordinate singularity at $R=S$ or $\alpha/R =1$ or $r=1$ now lies at the equator $\chi = \pi/2$. This indicates that the 3-space of the problem occupies one-half
of a 3-sphere of radius $S$. However such mathematical manipulation does not introduce any change of physics in the problem and one needs to be
concerned with the singularity at $R=\alpha$, or $R=S$, or, $r=1$ or $\chi = \pi/2$.

Let us suppose that somehow the density of the star would be reduced to zero everywhere. In such a case, the space time must become flat with a metric
\b
ds^2 = dt^2 - dR^2 - R^2 d\Omega^2
\e
From the viewpoint of metric(8) it would immediately be apparent that one should then have $e^\nu =1$ and $\alpha/R =0$. But from the viewpoint of
metric(14), it would appear that in such a case one would have $S= \infty$, $r=0$ and $R= S r \ge 0$. Why would we have $r=0$ in the latter case? This is so because the
general form of a space time with constant spatial curvature is
\b
ds^2 = e^\nu dt^2 -  S^2 \l( {dr^2\over 1 - K r^2} + r^2 ~d\Omega^2 \r)
\e
where $K= +1, 0, -1$ corresponds to closed, flat and open space time respectively. And once we presume $K=+1$ and adopt metric(14), we presume $\rho >0$
and then it becomes difficult to have a smooth transition to a $ \rho \to 0$ 
case. Hence, if one would try to arrive at a flat space ($K=0)$  by imagining that
it is due to a closed space of infinite radius, one would land up with the condition $r=0$. In fact, if in Eq.(4), had we taken, 
\b
T_0^0 = K \rho
\e
{\em instead of $\rho$, we would indeed have obtained Eq.(18) directly in lieu of  Eq.(14).  Then we would have found that the condition for flatness
is $K =0$}. But when we do take $T_0^0 = \rho$ and implicitly assume $\rho > 0$, we presume $K =1$ and exclude the possibility that one might have
$K=0$ too. In such a case, the condition for flatness would appear as $r=0$ as mentioned above.

\subsection {Cosmological Constant}
 
 It is well known that in order to obtain a static universe which would be closed and finite radius, Einstein
modified Eq.(1) into
\b
G_{ik}  + \Lambda g_{ik} = 8 \pi~ T_{ik}
\e
apparently implying that either $\Lambda$ is a fundamental constant like $8 \pi G/c^4$ or a basic scalar, like the Ricci scalar ${\cal R}$ appearing
within $G_{ik}$. 
 It follows that, now one has an ``effective''
density\cite{8}
\b
\rho \to \rho_e = \rho + {\Lambda \over 8 \pi}
\e
and an effective pressure
\b
p \to p_e = p - {\Lambda \over 8 \pi}
\e
So instead of Eq.(7), one now obtains
\b
e^{-\lambda} = 1 - {\alpha_e(R) \over R}
\e
where
\b
\alpha_e = \int_0^R 8 \pi \rho_e R^2 dR
\e
One also finds
\b
ds^2 = e^\nu dt^2 - {dR^2\over 1 - \alpha_e/R} - R^2 d\Omega^2
\e
And for a constant density case, one has 
\b
ds^2 = e^\nu dt^2 -   {dR^2\over 1 - (8 \pi  /3) \rho_e R^2 } + R^2 ~d\Omega^2 
\e
and
 \b
S = \sqrt{3\over 8 \pi \rho_e}
\e
 Eqs.(12-16) remains unaltered in the presence of $\Lambda$.
Thus now, if the space time has to be flat, in addition to $e^\nu =1$, one must have
\b
\rho_e \to 0
\e
 This shows that if the original definition of ``vacuum'' is $\rho =0$, as if, in the presence of a $\Lambda$, it  gets
modified to $\rho_e =0$. 
\section{ Acceleration Scalar and Singularity}
The fact that for a spherically symmetric static system, one  should have $\alpha/R <1$ has already been investigated\cite{11,12}.
The basic reason for this is not difficult to see. In general, static or non-static, for spherically symmetric space time occurrence of 
$\alpha(r,t)/R >$  corresponds to a formation of a ``trapped surface'' and the condition $\alpha(r,t)/R =1$ marks the formation of an ``apparent horizon''. 
Though for a non-static system, it might be possible to have trapped surfaces or horizons, for a static system they are not allowed. This is so because
once a trapped surface would be there, stellar matter would be inexorably pulled towards the central point of symmetry and thus matter would soon end up in a point singularity rather than as an extended static object. While this is definitely not allowed for a static star, this problem would
be much more severe for cosmology because ``center of symmetry lies everywhere''.

Following the case of a Schwarzschild black hole space time, generally, it is believed that this $R=\alpha$ singularity is a mere coordinate one
even in the presence of matter. But as we would see below, the coordinate independent {\em scalar acceleration} of the fluid would blow up unless
severe constraints are imposed on the fluid EOS.  

It may be recalled that Einstein was very much concerned about this $R=\alpha$ singularity and constructed a static model of a fluid where test particles are moving in randomly oriented circular orbits under their own gravitational field. While the radial stress of the fluid is zero, tangential
stresses are finite. This configuaration is known as ``Einstein Cluster'' and Einstein showed that, the speed of the orbiting particles would be equal to the speed of light $c$ at $R=\alpha$\cite{13}. Thus, he pleaded that there cannot be any $R=\alpha$ or $R < \alpha$ situation. Later it turned out that Einstein's cluster indeed corresponded to some well defined interior Schwarzschild solution studied by Florides\cite{14,15}.
And now Bohmer \& Lobo\cite{16} have shown that Einstein's intuition was correct, and the $R=\alpha$ singularity is indeed a {\em curvature singularity}.
This however does not at all mean that all $R=\alpha$ singularities are curvature singularities\cite{8}. In fact, in the present static case of isotropic pressure, it would be found
that they should be regions with zero curvature singularity to avoid an acceleration or pressure gradient singularities.

 For any static spherically symmetric fluid, one can easily find the acceleration\cite{10}
 \b
 a^i = u^k ~\nabla_k u^i
 \e
 In spherical symmetry, only one component of acceleration survives
 \b
 a^R = {e^{-\lambda} ~ \nu'\over 2}
 \e
 How to evaluate this $a^R$? Again, for the sake of easy understanding, we first do not consider any $\Lambda$ and write down the $_R^R$ component of Einstein equation\cite{2},
\b
8 \pi T_R^R = - 8 \pi p = - e^{-\lambda} \l( {\nu'\over R} + {1\over R^2} \r) + {1\over R^2}
\e
In view of Eq.(30), let us rewrite this equation as
\b
-  {e^{-\lambda}\over R} = {\alpha + 8 \pi p R^3\over R^3}
\e
or
\b
 a^R =  -{\alpha+ 8 \pi R^3 p\over 2 R^2  }
 \e
Since there is only one component of $a^i$, the {\em scalar acceleration} becomes 
 \b
 a = \sqrt{-a^i a_i} = a^R \sqrt{|g_{RR}|} ={e^{-\lambda/2} |\nu'| \over 2}
 \e
Using Eq.(33), we obtain
 \b
 a =  {\alpha + 8 \pi R^3 p\over 2 R^2 \sqrt{1 - \alpha/R} }
 \e
 If $\Lambda$ would be included, then the acceleration scalar would be given by
\b
 a =  {\alpha_e + 8 \pi R^3 p_e\over 2 R^2 \sqrt{1 - \alpha_e/R} }
 \e
It is now clear that if a static fluid would tread upon the $R=\alpha_e$ singularity, its acceleration could be infinite
and it cannot be at rest. And this is the physical reason that for a static fluid  one must have $R > \alpha_e$ everywhere. And
in case, the manifold would cover $R=\alpha_e$, one must satisfy 
\b
\alpha_e + 8 \pi R^3  p_e =0; \qquad R=\alpha_e
\e
in an attempt to keep $a$ regular.
Using $R=\alpha_e$ in the foregoing equation, we have
\b
\alpha_e( 1 + 8 \pi \alpha_e^2 p_e) =0
\e
i.e., one must have either
\b
\alpha_e =0
\e
or,
\b
p_e (R=\alpha_e) = -{1 \over 8 \alpha_e^2}
\e
or both of the above two conditions. If we assume that the minimum value of $p$ is zero and it cannot be negative, we will have
$p_e = - \Lambda/8 \pi$. Then Eq.(40) would yield
\b
R =S = \sqrt {1\over \Lambda}
\e
But one is still not sure whether additional condition such as Eq.(39) is needed to really ensure that $a$ is indeed finite at $R=\alpha_e$.
 For a fluid with $\rho_e >0$, the safest way to avoid the $R=\alpha_e$ singularity will simply be to ensure that
 $ R > \alpha_e$. The fluid can do so by choosing an appropriate density profile and by ensuring that its outer boundary
\b
R_b > \alpha_e
\e
For a constant density star or the ESU, we have
\b
a = \kappa {R (\rho_e + 3 p_e) \over \sqrt{1 - 2 \kappa \rho_e R^2}} 
\e
where $\kappa = 4\pi/3$. In terms of $r$, we obtain
\b
a = \kappa S { r (\rho_e + 3 p_e )\over \sqrt{1 - r^2}}
\e

It is clear that the sufficient condition for avoiding the $R=\alpha_e$ singularity in this case is
\b
R_b > \alpha; \qquad R <S
\e
In terms of density, this means
\b
\rho_e > { 3\over 8 \pi R_b^2}
\e
Thus, if one would imagine a region with $\rho =0$, i.e., $\rho_e = \Lambda/8\pi$, one should restrict the radius of this vacuum as
\b
R_b < \sqrt{ 3\over \Lambda}; 
\e
And in case, one would have $\rho = \Lambda/4\pi$, then the above restriction would become
\b
R_b < \sqrt{ 1\over \Lambda}
\e
and which cannot be satisfied\cite{8}. Thus for a constant density star with $R_b = \sqrt{1/\Lambda}$, {\em one must have} $\rho >  \Lambda/ 4\pi$. And in case this condition would be
violated, one must critically analyze the additional constraint on the EOS which would prevent $p_e'$ from blowing up at $R=\alpha_e$. It may be seen that the $\rho = \Lambda /4 \pi$ EOS corresponds to
\b
\rho_e + 3 p_e =0
\e
if $p=0$.

\section{ TOV Equation}
Recall that local energy momentum conservation equation ${T_i^k}_{; k} =0$ immediately leads to\cite{2}
\b
\nu' = {- 2p'\over p +\rho}
\e
Further if  we again consider the $_R^R$ component of Einstein Eq. and combine it with Eq.(50), we will obtain the TOV equation for hydrostatic balance for any self-gravitating static fluid:
\b
p' = - {(\rho +p) (\alpha + 8 \pi R^3 p)\over 2 R^2 (1 - \alpha/R)}
\e
Since the effect of $\Lambda$ gets included if one replaces $p$ and $\rho$ by their ``effective'' values\cite{8}
\b
p' = - {(\rho_e +p_e) (\alpha_e + 8 \pi R^3 p_e)\over 2 R^2 (1 - \alpha_e/R)}
\e
Further since $p' = p'_e$, in a complete symmetric manner, we rewrite the above equation as
\b
p'_e = - {(\rho_e +p_e) (\alpha_e + 8 \pi R^3 p_e)\over 2 R^2 (1 - \alpha_e/R)}
\e
This equation strongly suggests that if the definition of a ``dust'' in the absence of $\Lambda$ is $p=0$, as if, in the presence of $\Lambda$,
dust EOS should be $p_e =0$. Similarly, if the original EOS of ``vacuum'' is $\rho=0$, in the presence of $\Lambda$, the vacuum EOS is $\rho_e=0$,
a hint we have already found.

Note, even now, there is no need to interpret  $\alpha$ in terms of any {\em exterior boundary condition}, and thus TOV equation
is valid in any spherically symmetric static GR problem including ESU. By using Eq.(36), it is interesting to rewrite the TOV Eq. in terms of the acceleration scalar
\b
p'_e = - { a (\rho_e +p_e) \over \sqrt{ 1 - \alpha_e/R}}
\e
Clearly, there is a singularity in the denominator of TOV Eq. at $R=\alpha_e$. If we write
\b
x =\sqrt{1 - \alpha_e/R}
\e
it is seen that while the singularity in acceleration $a \sim x^{-1}$, for the pressure gradient it is much stronger $p' \sim x^{-2}$. And  if the fluid would cover the $R=\alpha_e$ singularity, regularity of $a$ may not be sufficient to ensure regularity of $p'$. In fact Eq.(54) suggests that
one might require the additional constraint
\b
\rho_e + p_e =0; \qquad R=S
\e
to tame the $p'$ singularity at $R=S$.

 Now let us use the condition $\rho_e = uniform$ which is equally valid for
the interior solution of a constant density star as well as the ESU:
\b
p' = -  \kappa R{(\rho_e +p_e) (\rho_e + 3 p_e)\over  1 -  2\kappa \rho_e R^2}
\e
In terms of the normalized coordinate $r$, one finds
\b
p' = -  \kappa  S {r(\rho_e +p_e) (\rho_e + 3 p_e)\over  1 -  r^2}
\e
And in term  of $a$, we have
\b
p' = -   {a(\rho_e +p_e) \over  \sqrt{1 -  2\kappa \rho_e R^2}}
\e
and
\b
p' = -     {a (\rho_e +p_e) \over  \sqrt{1 -  r^2}}
\e
\section {Constant Density Star}

For the  interior solution of a constant density relativistic star one, one finds\cite{8}
\b
e^\nu = \l( {p^c + \rho\over p + \rho} \r)^2
\e
where $p_c$ is the {\em central} pressure. This relation is interesting because it does not involve any exterior bounday condition. Also note that
since
\b
p + \rho = p_e + \rho_e
\e
we can rewrite Eq.(61) as
 \b
e^\nu = \l( {p_e^c + \rho_e\over p_e + \rho_e} \r)^2
\e
Since for a star, there must be a pressure gradient, $p^c \neq p(R)$, and  one really cannot reduce $e^\nu (R)$ to a constant value independent
of $R$. 
But if, from the mathematical viewpoint, one would still demand that it should be possible to set up a time orthogonal Gaussian coordinate
system where $dt^2 = d\tau^2$, one would unknowingly kill the pressure gradient and set $p^c = p$ in Eq.(61). In the absence of $\Lambda$, for a static configuration,  $\rho=0$ if $p=0$, and the star would vanish under the assumption of $e^\nu =1$!

When $\Lambda$ is present, the expression for pressure for a constant density star is\cite{8}
\b
p(R) = \rho { \Lambda/4 \pi -1 + C \cos\chi \over 3 - \cos \chi}
\e
where
\b
C = { 3 p^c + \rho - \Lambda/4 \pi \over p^c + \rho} = { 3p^c_e + \rho_e\over p^c_e + \rho_e}
\e

Now it transpires that even when $p=0$ to honor $e^\nu =1$, there would still be a finite density $\rho = \Lambda/ 4 \pi$. But one attains
this finite density at a huge cost because it now turns out that the {\em boundary of the star merges with the coordinate singularity $R_b= S$
where the metric behaves badly}. For a real star, this singularity can always be avoided by allowing $e^\nu >1$ which in turn means by having
$\rho >  \Lambda/4\pi$. Thus a ``world time cannot be introduced for a general relativistic star having a finite density.

\subsection{ Verification From TOV Equation}
Although the case of constant density stars in the presence of cosmological constants has been studied in great detail\cite{8}, to our knowledge,
nobody actually verified the validity of the solutions in case one would tread on the coordinate singularity at $R=S$. Since the denominator of Eq.(53)
becomes zero at $R=S$, in order that $p'$ does not blow up there, one must have either 
\b
\rho_e + p_e =0; \qquad R=S
\e
or
\b
\rho_e + 3p_e =0; \qquad R=S
\e
or both
\b
\rho_e + p_e = \rho_e + 3p_e =0; \qquad R=S
\e
If the last constraint would be satisfied, one would immediately obtain $\rho_e =p_e =0$. However, it is possible that Eq.(68) is not satisfied and only one of the less rigourous conditions(66) or (67) is satisfied. In particular, to avoid occurrence of negative pressure, let us assume that only Eq.(67) is satisfied at $R=S$. Since
 $p'$ attains a $0/0$ form at $R=S$, let us study the nature of $p'$ at this singularity by using
L' Hospital rule. For this, let us first write
\b
p_e' = {f(R) \over g(R)}
\e
where
\b
f(R) = \kappa ~R (\rho_e +p_e) (\rho_e + 3p_e)
\e
and
\b
g(R) = 2 \kappa \rho_e R^2 -1
\e
Since $\rho'=0$, we find that
\b
f' = \kappa (\rho_e +p_e) (\rho_e + 3p_e) + \kappa  R  p'_e [3(\rho_e + p_e) +  (\rho_e + 3p_e)]
\e
Since, we have already considered $\rho_e + 3p_e =0$, we obtain a reduced expression for 
\b
f' = 3 \kappa R p_e' (\rho_e +p_e)
\e
On the other hand,
\b
g' = 4 \kappa \rho_e R
\e
So that
\b
{f'\over g'} = 3 p_e' {\rho_e + p \over 4\rho_e}
\e
And at, the singularity, $R= S$, by l' Hospital rule, we obtain
\b
p_e' = \lim_{R\to S} {f'\over g'}
\e

From Eqs.(75) and (76) we obtain the required condition:
\b
 3 (\rho_e +p_e) = 4 \rho_e
\e
i.e.,
\b
\rho_e (R=s) = 3 p_e (R=S)
\e
which looks like the EOS of incoherent radiation! By combining Eqs.(67) and (78), it becomes clear that in order that $p'$ does not blow up at
the ``coordinate singularity'' at $R=S$, the constant density star must have
\b
\rho_e (R=S) = p_e (R=S) = 0
\e
Thus instead of $p= 0$ we would obtain $p_e = 0$ at $R=S$. This means that if indeed $\Lambda >0$, the solution must avoid $R=S$ singularity.
More importantly, since $\rho_e = constant$, we find that, if the solution would indeed extend upto $R=S$, we must have $\rho_e =0$. Thus all
finite density stars must avoid the $\alpha_e/R =1$ or $r=1$ singularity.

\section{Static Universe}
In cosmological case, there must be a universal time which would be the proper time of all fundamental observers,
i.e., $dt^2 = d\tau^2$. This demands that  one must be able to set
$e^\nu =1$ so that, the Eq.(14)  would become the ESU metric
\b
ds^2 = dt^2 -  S^2 \l( {dr^2\over 1 -  r^2} + r^2 ~d\Omega^2 \r)
\e
Note that the value of $S$ for ESU is still given by Eq.(27)\cite{10} and thus one is justified in deducing ESU metric indeed by setting $e^\nu =1$. Since the  Einstein equation tells that the metric is essentially determined by $T_{ik}$, such an important change of setting $e^\nu =1$
 on the LHS of this equation must be endorsed by the RHS, i.e., by the admissible forms of fluid EOS. What are those conditions? To explore
 them, we note from Eqs.(50) and (51) that the most general form of static, spherically symmetric Einstein Eq. yields
 \b
 \nu' = - {\alpha_e + 8 \pi R^3 p_e\over R^2 (1- \alpha_e/R)}
 \e
In constant density case, this condition becomes
\b
 \nu' = - 2\kappa  {R(\rho_e + 3 p_e)\over   1- 2 \kappa \rho_e R^2}
 \e
 And in term of $r$, this condition becomes
 \b
 \nu' = - 2\kappa  S {r (\rho_e + 3 p_e) \over   1- r^2}
 \e
In order that $e^\nu =1$, one must have $\nu' =0$, and from the foregoing equation, it is immediately clear that
then we must have atleast 
\b
\rho_e + 3p_e =0
\e
In any case, we must have $p_e' =0$ everywhere including $R=S$ or $r=1$. Then if  we follow the L' Hospital treatment of the previous section,
we would find, we must have
\b
\rho_e =0
\e
everywhere because $\rho_e =constant$. Although, we have already obtained this important result, it would be interesting to obtain this
result from somewhat different routes.

First consider Eq.(59) and write
\b
f(R) = a (\rho_e + p_e)
\e
and
\b
g(R) = - \sqrt{ 1 - 2 \kappa \rho_e R^2}
\e
so that
\b
f' = p_e' a
\e
and
\b
g' = 2 \kappa \rho_e (1 - 2 \kappa \rho_e R^2)^{-1/2}
\e
Now applying L'Hospital rule at $R=S$, we find
\b
p' = p' {a\sqrt{1- 2\kappa \rho_e R^2} \over 2 \kappa \rho_e R}
\e
Using Eq.(43) in the above Eq., we find
\b
{\rho_e + 3 p_e\over 2 \rho_e} =1
\e
implying
\b
\rho_e + p_e =0
\e
i.e., both the factors in the numerator of Eq.(58) should vanish to ensure regularity of $p'$ at $R=S$. It may be also of some interest to
extend this study by directly considering the radial variable as $r$ rather than $R$. Note that
\b
p_e' = {1\over S} {dp_e \over dr}
\e
Now if we denote differentiation by $r$ with a prime, we rewrite Eq.(60) as
\b
p' = -{S a  (\rho_e +p_e)\over \sqrt{1-r^2}}
\e
so that
\b
f(r) = S a (\rho_e + p_e)
\e
and
\b
g(r) = - \sqrt{1 -r^2}
\e
Then we have
\b
f' = S a p'
\e
and
\b
g' = {r \over \sqrt{1-r^2}}
\e
Again applying L'Hospital rule for the limit of $p'$ at $R=S$, we find
\b
{S a \sqrt{1-r^2}\over r} =1
\e
Inserting the expression for $a$ from Eq.(44), we obtain the interesting relation
\b
\kappa S^2 (\rho_e + 3 p_e) =1
\e
or
\b
S = \sqrt { 3 \over 4 \pi (\rho_e + 3 p_e)}
\e
Comparison of Eqs.(27) and (101) would again convey one of the hidden mesages for the ESU as well as for a constant density star which is attemting to
suppress its pressure gradient, namely $3 p_e = \rho_e$.
The sum and substance of the entire exercise is that for the ESU, in oder that $p_e'$ is indeed zero everywhere, one must have
\b
\rho_e = 3 p_e =0
\e
 But unlike the constant density stellar
case {\em now we cannot escape confronting this singularity by demanding that $\rho >\Lambda/4\pi$}. And hence we must accept the fact that for the ESU fluid $\rho_e = p_e=0$ everywhere. This means that
\b
\rho + \Lambda/ 4\pi =0; \qquad  \rho = - 4 \Lambda/4 \pi
\e
atleast for the ESU. The weak energy condition demands that $\rho \ge 0$ and thus we find that $\Lambda =0$!

\section{Einstein's Solution}
If one would directly use Eq.(80) into Einstein Eq.(20), one would be led to\cite{5}
\b
4 \pi (\rho + 3 p) = \Lambda
\e
and
\b
4 \pi (\rho + p) = {1\over S^2}
\e
Einstein chose, $p=0$ EOS which led to
\b
{1\over S^2} = \Lambda; \qquad S=\sqrt{1\over \Lambda}
\e
Note that Eqs.(104) and (105) may be rewritten as
\b
4 \pi (\rho_e + 3 p_e) =0
\e
and
\b
4 \pi (\rho_e + p_e) ={1\over S^2}
\e
But, if we recall, Eq.(92), actually, $\rho_e + p_e =0$ so that $S=\infty$. Further Eqs.(100-101) showed that
\b
4 \pi (\rho_e + 3 p_e) = {3 \over S^2}
\e
And a comparison of Eqs.(108) and (109) again shows that, $S=\infty$ and $\rho_e =0$. As mentioned before, for weak energy condition,
this would mean $\rho = \Lambda =0$. Note that  it is indeed  possible to have  a situation where $\rho_{rest}/e \gg 1$, where $e$ is the proper internal energy density. In such a case,
one might {\em approximately} write $p \approx 0$. But this does not mean that pressure is strictly zero. A {\em strict} $p=0$ EOS is possible only
when $\rho =0$. In such a case, Eq.(104) would again yield $\Lambda =0$.
 
 If we ignore such physical and mathematical regularity or self-consistency considerations, Einstein's solution would suggest $\Lambda = {4 \pi G \rho/ c^2}$.
 
 Further, if one would assume that (i) ESU correctly describes the real universe and (ii) the true mean density of the real universe in its totality (about which
 we may never have absolute knowledge)  is equal to the mean density of the {\em patch} of the observed universe, i.e.,
 \b
 \rho_{patch} = \rho_{true} = \rho \sim 10^{-31} ~{\rm g ~cm^{-3}}
 \e
 one would obtain\cite{3}
 \b
 \Lambda \sim 10^{-58} ~ {\rm cm^2}
 \e
 Now ponder over the fact that until the advent of big optical telescopes in 1920 or so, for hundreds of years, most of the astronomers thought
 that total universe was nothing but our milkyway galaxy. And if would have been possible to measure the mean density of the Milkyway, one might have concluded that
 \b
 \rho_{patch} = \rho_{true} = \rho \sim 10^{-13} ~{\rm g ~cm^{-3}}
 \e
 and accordingly
 \b
 \Lambda \sim 10^{-40} ~ {\rm cm^2}~~before ~1920
 \e
 As late as 1970, we hardly had any idea that the galaxy clusters and superclusters are actially distributed as ``filaments'' and ``walls''
 around huge voids whose dimensions could be as large as $\sim 280$ Mpc\cite{17}. The recent Sloan Digital Sky Survey has revealed structures of dimension $\sim 500$ Mpc\cite{18}. Thus, we cannot rule out the possibility that that eventually, it might be
 found that  the entire patch of presently observed universe also lies on the wall of a larger void. If the mean density of observed universe would be revised by
 future observations, would we again revise the value of the {\em fundamental constant} $\Lambda$?
 
 Many cosmologists believe that within the patch of the observed universe, galaxies are distributed in fractal like pattern with length scale
 $l \sim 100$ Mpc and a fractal dimension $D\sim 2$ \cite{19, 20,21}. One may recall here a succint comment by Wu, Lahav \& Rees\cite{22}
 
 ``The universe is inhomogeneous - and essentially fractal - on the scale of galaxies and clusters of galaxies, but most cosmologists believe that on larger scales it becomes isotropic and homogeneous.''

In a fractal distribution, the number of galaxies increases with $R$ as
 \b
 N(< R) \sim R^D
 \e
 and the number density of galaxies varies as
 \b
 n(<R) \sim R^{D-3}
 \e
 Thus, in principle, for a fractal distribution over sufficiently large scale, one may have $n\to 0$, $\rho \to 0$.  Most of the fractal patterns however tend to have a preferred center and thus not suitable for an eventually
 homogeneous and isotropic universe. But there could be some fractal structures like {\em Levy Dust} which may not have any preferred center. 
 One may have a mean $\rho =0$ also
 for an appropriate infinite {\em multi universe}\cite{23, 24}.
 Suppose the entire observed universe where we live is only a speckle in a multi universe comprising infinite
 such speckles. If these speckles or ``cosmic atoms'' are separated from one another by infinite voids, then the mean density of the infinitely
 diluted multi universe would be zero.  
 
 At the cost of sounding repetitive, let us again raise the old question:  if $\Lambda$ would indeed be due to some quantum mechanical effect,  why the value of $\Lambda$ obtained  under the {\em assumption} of $\rho_{patch} = \rho_{true}$   falls short of the theoretical value by an approximate factor of $10^{120}$?
 And if the various field modes would cancel one another to generate a small $\Lambda$, why do they not cancel exactly to result $\Lambda =0$?
 In fact there are some theoretical estimates by which $\Lambda =0$\cite{25}.
 
 In physics, we have various fundamental constants like $G$, $c$, $h$, $m_e$ etc. But in no case there is any evidence that the values of such constants
 directly depend of ambient factors like mean density of universe. In fact, all truly fundamental factors allow themselves to be determined
 with high precision by judicious combination of theory and experiments. The only exception here is the supposed $\Lambda$! 
 
 Thus if one would indeed need to invoke a source of ``repulsive gravity'' in some model, the source cannot be a fundamental time independent constant ($\Lambda$)
 sitting on the LHS of Einstein equation (20). On the other hand, such repulsive gravity could be due to a ``dark energy'' fluid represented by an appropriate $T_{ik}$ sitting on the RHS of Einstein equation. In such a case, in a non-static case, the energy density of the dark energy fluid also
 must be non-static and except for a fiducial initial value, such an energy density may not considered as any ``fundamental parameter''.
 \subsection{ Faintness of Distant Supernovae}
 Over the past decade, cosmologists have gathered evidences that  the luminosities of distant Type 1a supernove are lower than what is expected in the standard FRW model.  Here, 
 it is assumed that the Type Ia supernovae luminosities are indeed ``standard candles'' despite the likely evolution of luminosities due to
 likely cosmic evolution. Even if the Type 1a supernovae would be standard candles without appreciable cosmic evolution, there is some chance
 that their faintness could be due to opacities of distant inter galactic medium. For example, it has been pointed out that Lyman Alpha clouds might
 introduce some non-transparency for optical emission from distant universe\cite{26}. It has been also argued that the atmosphere of planets could be additional
 sources of opacities for very distant supernovae lights\cite{27}.
 
  If we ignore this possibility of dimming of distant supernovae due to some intergalactic opacity, for the FRW model, one needs to postulate an
  ``accelerating universe'' where the acceleration is driven by some source of repulsive gravity. But as discussed in the previous subsection, this
  repulsive gravity should be due to a dark energy fluid whose energy density is decayed from the quantum gravity value by a factor of almost $10^{120}$. And this repulsive source of gravity should appear from $T_{ik}$ rather than as a fundamental constant from the LHS of Einstein equation.

 Also, one should leave here an open window for some hitherto unknown physics or surprise for understanding this new phenomenon.  For instance even now, there is no firm evidence that Copernican principle, the basis of the standard cosmology is indeed valid\cite{28}. Also, consider the fact that we are able to ``see'' galaxies and measure their redshifts because they are radiating. But in the standard cosmologies, galaxies
 are considered as non-radiating neutral ``dust'' particles.

 \section{ Summary and Discussions}
 Both a spherically symmetric static star and the ESU result from the same spherically symmetrical form of Einstein equations. If $_R^R$ component
 of the Einstein equation would be studied, one would obtain, the expression for acceleration scalar and condition for hydrostatic balance in
 both the cases. In fact, Bohmer\cite{8}  did obtain {\em one of the constraints to be followed by the ESU fluid by demanding that the pressure gradient $p'$ must vanish for the ESU}: $\rho_e + 3 p_e =0$.  We found that the singularity appearing in the $(R, R)$ component of the metric tensor in all spherically symmetric static configurations do propage in the expressions for {\em acceleration scalar} $a$ and pressure gradient $p'$. While, for the former, the nature of singularity
 is $a \sim x^{-1}$, for the latter it is much stronger $p' \sim x^{-2}$. In fact, the requirement $\rho_e + 3 p_e =0$ is the basic condition for ensuring
 that the $a$ does not blow up at $x=0$ or $r=1$. This basic condition however need not be the sufficient condition for ensuring that $a$ is finite at
 $x=0$. Further, since $p'$ tends to blow up much faster at $x=0$, one must require an additional condition to ensure regularity of $p'$. We found that the latter requirement is $\rho_e = 3 p_e$. This means that if the constant density static star would have $R_b = S$, one must have $\rho_e =0$.
 But one can always avoid this singularity by ensuring that $\rho > \Lambda/ 4\pi$. However, previous detail  studies of perfect fluid spheres
 were primarily intended for
 isolated systems and not for
 the universe, and to our knowledge, nobody tried to study the remedy for either the $\sim x^{-1}$ or $\sim x^{-2}$ singularities at $x=0$.
 
 Our result that whether it is the ESU or a star with $R_b =S$ should have $\rho_e =0$ to tame the $p'\sim x^{-2}$ singularity can be obtained probably from other directions as well.
 Consider the interior metric(7) where $t$ is the time measured by a clock at a fixed $R$. In the exterior vacuum Schwarzschild metric
 \b
 ds^2 = (1 - \alpha_e/R) dt^2 - {dR^2 \over 1 - \alpha_e/R} - R^2 d\Omega^2
 \e
 too, time is measured by a clock resting at $R=R$. Suppose we match the interior and exterior solutions at $R=R_b$. If such a matching is possible,
 we would have
 \b
 e^{\nu_b} = 1 - \alpha_e/R
 \e
 In particular, consider the constant density solution\cite{8} where
 \b
 e^{\nu_b} = {p^c + \rho\over \rho}
 \e
 In order that $e^{\nu_b} \to 1$, it is clear that, one must have
 \b
 \alpha_e/R_b = {8 \pi \over 3} \rho_e R_b^2 \to 0
 \e
 which implies
 \b
 \rho_e \to 0
 \e
 It is also seen from Eq.(118) that $e^{\nu_b} \to 1$, only if $p_c \to 0$. It so happens that, one fails to match the interior solution with the exterior vacuum solution(7) at $\rho = \Lambda/4 \pi$. As we have found, one fails to do so because $g_{RR}$ blows up in Eqs.(10) and (116) in this case.
 
\subsection{Buchdahl Inequality}
It is well known that any spherically symmetric static configuration, homogeneous or inhomogeneous, satisfies a constraint {\em much stronger
than} the $R_b> \alpha$ or $\alpha < R_b$ constraint. In the absence of a $\Lambda$, this is known as Buchdahl inequality\cite{1, 26}:
\b
{\alpha\over R_b} < {8\over 9}
\e
If one would have $\alpha = (8/9) R_b$, central pressure would blow up (assuming $\rho >0$).  For a uniform density star this borderline would correspond to
\b
r = 8/9
\e
But suppose there is a solution which appears to violate
this constraint, i.e., $r \ge 8/9$. Then how to ensure that central pressure is still finite? The only solution out of this dilemma
would be to set $\alpha =0$ or $\rho =0$. From this more stringent condition, it should be clear why Einstein's original static universe
must have $\rho =0$.

Does the fate of the ESU improve after the incorporation of a positive $\Lambda$? From Eq.(3.35) of ref.\cite{8}, it turns out that now the modified
Buchadhal constraint is
\b
{\alpha_e \over R} < {2\over 9} \l( 4 - {\Lambda \over 4 \pi \rho}\r)
\e
Thus with a positive $\Lambda$, this constraint gets even tighter. With $\rho = \Lambda/ 4\pi$, one obtains
\b
{\alpha_e \over R} < {2\over 3} 
\e
which means, for a constant density star, one must have
\b
r < 2/3
\e
to ensure that pressure does not blow up at the center $r=0$! For the ESU, this would mean blowing up of pressure everywhere because ``center would lie everywhere''. And the only solution to get rid of this problem is is accept the fact that $\alpha_e =0$ for ESU.

\subsection{Mass Function}
Since for a fluid in an asympotictially flat space time, one can define a ``mass function'' in terms of which acceleration scalar
\b
 a =  {M + 4 \pi R^3 p\over  R^2 \sqrt{1 - 2 M/R} }
 \e
we can now identify $\alpha(R)$ in Eq.(36) as twice the quasilocal mass-energy of the fluid
\b
\alpha(R) = 2 M(R)
\e
In the presence of a $\Lambda$, the quasilocal mass will be
 $M_e = \alpha_e/2$. 
 
 In general relativity, the gravitational mass of a stationary system is\cite{4}
 \b
 M = \int (T_0^0 + t_0^0) ~\sqrt{-g} ~d^3 x
 \e
 where $t_i^k$ is energy momentum pseudo tensor and $g$ is the determinant of the metric tensor. One can work out $t_0^0$ from the metric $g_{ik}$ of the universe.
 Probably starting from Rosen\cite{27}, many authors have worked out the mass energy of a closed universe and all of them have concluded that for a closed
 universe $M=0$\cite{28,29,30,31,32}. This too would tell that, in the absence of $\Lambda$, $\rho =0$, and, in general $\rho_e =0$. For the ESU, it is possible to confirm this result irrespective of the value of $t_0^0$. In all coordinate systems, for a static fluid, one obtains\cite{28, 29, 30}
 \b
 M_e=\int (\rho + 3p -\Lambda/4 \pi) ~\sqrt{-g} ~d^3 x =\int (\rho_e + 3p_e) ~\sqrt{-g} ~d^3 x 
 \e
 In view of Eq.(84) we then directly obtain
 \b
 M_e =0
 \e
Then from Eq.(24),  we obtain $\rho_e =0$
\subsection{Poisson Equation}
It is known that the RHS of Poission equation indicates the source of gravity. For a spherically symmetric static fluid the Poission's equation is
\b
\nabla^2 \sqrt{g_{00}} = 4\pi \sqrt{g_{00}} (\rho + 3p - \Lambda/4 \pi) =4\pi \sqrt{g_{00}} (\rho_e + 3p_e)
\e
This shows that the source density of gravity is
\b
\rho_g = \sqrt{g_{00}} (\rho_e + 3p_e)
\e
And when $g_{00} =1$, one will obtain
\b
\rho_g =  (\rho_e + 3p_e)
\e
 In either case, the RHS of Eq.(131) is zero when $\rho_e + 3 p_e =0$, which is the case for both a constant density star ({\em if it would be extended
 up to $R=S$}) or the ESU.  In all such cases, one must necessarily have $g_{00} = e^\nu = constant$. And since source of gravity is zero, the space time
 must be flat. And a space time is flat when $\rho_e =0$, as obtained by us. Alternatively, if one would first focus attention on the LHS of Eq.(131),
 it seems that if one would set $e^\nu = constant$, the source of gravity will vanish, again in which case, one should have $\rho_e =0$.
\subsection{A Simple Reason}
For a constant density case, we may rearrange the TOV equation(58) as
\b
{-p'\over  (\rho_e + p_e) (\rho_e + 3 p_e)} = { \kappa S~ r \over 1 - r^2}
\e
We may also rewrite the acceleration equation(44) as
\b
{a \over \rho_e + 3 p_e} = { \kappa S~r \over \sqrt{1-r^2}}
\e
The RHS of Eqs.(134) and (135) obviously depends on $r$ if $r$ would indeed be a free parameter. For a constant density star, the LHS of the same equations too
must depend on $r$. And they do depend on $r$ because $p_e=p_e(r)$ even though $\rho_e =constant$. But suppose we would like to freeze the
$r$ dependence of $p$ to make $e^\nu =1$. Then the value of $r$ on the RHS of Eqs.(134) and (135) too must be frozen. And since $R= S r \ge 0$ during the freezing process, the only solution here would be to adopt
\b
S = \infty; \qquad r=0; \qquad R\ge 0
\e
We have already discussed that occurrence of Eq.(136) actually signifies that the spatial section is flat: $K =0$. This would have been more transparent had we taken
\b
T_1^1 = T_2^2=T_3^3 = - K p
\e
alongwith Eq.(19), i.e., had we written
\b
T_{ik} = K [(\rho+p) u_i u_k +  p g_{ik}]
\e
instead of Eq.(4).  Further had we also taken the cosmological constant as $K \Lambda$ instead of $\Lambda$, Eqs.(134) and (135) would have appeared as
\b
{-p'\over  (\rho_e + p_e) (\rho_e + 3 p_e)} = { K \kappa S~ r \over 1 - Kr^2}
\e
and
\b
{a \over \rho_e + 3 p_e} = { K\kappa S~r \over \sqrt{1-Kr^2}}
\e
respectively.
In such a case, it would have been seen immediately that the requirement that the RHS of Eqs.(139) and (140) are independent of $r$, one must have is $K=0$.
And since we have already set
$e^\nu =1$, it means, the space time must be flat to ensure that the RHS of Eqs.(134) \& (135) or (139) \& (140) are indeed independent of $r$. Hence one must have $K\rho_e =0$
in such a case. So will be the case for ESU.
\section{Conclusions}
Cosmological constant was introduced as a fundamental constant on the LHS of the Einstein equation. And we found that, if $\Lambda$ would indeed be considered as   a time independent fundamental constant, its value is exactly zero rather than lower than the quantum gravity value by a factor of $\sim 10^{120}$. This, we believe, has been shown for the first time and is an important result in its own right. Thus the Einstein equation should be Eq.(1): $G_{ik}  = 8 \pi~ T_{ik}$ rather than Eq.(20). However, in principle, one is free to choose any kind of source of gravity, i.e., $T_{ik}$. In particular, one may choose
$T_{ik}$ as a blend of a normal fluid with positive pressure and a ``dark energy fluid'' with negative pressure. The latter fluid would
then act as a  source of ``repulsive gravity''. The most popular model to explaining the extra faintness of distant supernova invokes such quintessence
or dark energy fluid. Since in a non-static  universe model, all components of $T_{ik}$ including the energy density of dark energy fluid would be time 
 dependent, it would be easier to understand why the required present day density of this dark energy fluid is $\sim 10^{120}$ times lower than the
quantum gravity value. On the other hand, had this ``repulsive gravity'' been due to a time independent $\Lambda$, this task would have been much more difficult.




\begin{thebibliography}{99}
\bibitem{1}  S. Weinberg, {\it Gravitation and Cosmology: Principles and
Applications of General Theory of Relativity},  (John Wiley, New York,
1972)
\bibitem{2} L.D. Landau \& E.M. Lifshitz, {\it Classical Theory of Fields}, (Pergamon, Oxford 1962)

\bibitem{3} H. Weyl, Phys. Z., 20, 31 (1919)
\bibitem{4} R.C. Tolman, Phys. Rev., 55, 364 (1939)
\bibitem{5} R.C. Tolman, {\it Relativity, Thermodynamics \& Cosmology}, (Clarendron Press, Oxford 1950)
\bibitem{6} H. Nariai, Gen. Rel. Gravit., 31, 963 (1999)
\bibitem{7} Z. Stuchlik, Acta Phys. Slovaca, 50, 219 (2000)
\bibitem{8} C.G. Bohmer, Diploma Thesis, Potsdam Univ. (2002), gr-qc/0308057
\bibitem{9} C.G. Bohmer, Gen. Rel. Grav., 36, 1039 (2004)
\bibitem{10} C.G. Bohmer \& G. Fodor, Phys. Rev. D., 77, 064008 (2008)
\bibitem{11} A.D. Rendall \& B.G. Schmidt, Class. Q. Grav., 8, 985 (1999)
\bibitem{12} T.W. Baumgarte \& A.D. Rendall, Class. Q. Grav., 10, 327 (1993)
\bibitem{13} A. Einstein, Ann. Math., 40, 922 (1939)
\bibitem{14} P.S. Florides, Proc. R. Soc. Lond., A 337, 529 (1974)
\bibitem{15} N.K. Kofinti, Gen. Rel. Grav., 17, 245 (1985)
\bibitem{16} C.G. Bohmer \& F.S.N. Lobo, (2007), Int. J. Mod. Phys., D17, 897 (2008), gr-qc/0703024
\bibitem{17} L. Rudnick, S. Brown \& L.R. Williams, Astrophys. J., 671, 40 (2007)
\bibitem{18} J.R. Gott. III, M. Juric, I.D. Schlegel et al., Astrophys. J., 624, 463 (2005)

\bibitem{19} Y. Baryshev \& P. Teerikopri, {\it Discovery of Cosmic Fractals} (World Scientific, 2002)
\bibitem{20} B.B. Mandelbrot, {\it The Fractal Geometry of Nature}, (Freeman, NY 1982)
\bibitem{21} K. Wu, O. Lahav \& M. Rees, Nature, 397, 225 (1999)
\bibitem{22} V. Martinez, Science, 285, 445 (1999)
\bibitem{23} G.R.F. Ellis, U. Kirchner, \& W.R. Stoeger, MNRAS, 347, 921 (2004), astro-ph/0305292
\bibitem{24} A. Barrau, Cern Courier, Nov 20 issue (1997) 
\bibitem{25} S. Coleman, Nucl. Phys. B, 310, 643 (1998)

\bibitem{26} R.E. Schild \& M. Dekker, AN, 327(7), 729 (2006), astro-ph/0512236
\bibitem{27} C.H. Gibson \& R.E. Schild, Proc. ``Problems in Practical Cosmology'', p. 232 (St. Petersburg, 2008, ISBN: 978-5-902632-06-1), (astro-ph), arXiv:0803.4293
\bibitem{28} R.R. Caldwell \& A. Stebbins, Phys. Rev. Lett., 100, 191302 (2008)
\bibitem{29} H.A. Buchdahl, Phys. Rev., 116, 1027 (1959)
\bibitem{30} N.  Rosen,  Gen. Rel. Grav., 26, 31 (1994)
\bibitem{31} F.I. Cooperstock, Gen. Rel. Grav., 26, 323 (1994)
\bibitem{32} V.B. Johri, D. Kalligas, G.P. Singh, \& C.W.F. Everitt, Gen. Rel. Grav., 27, 313 (1995)
\bibitem{33} N. Banerjee, N \& S. Sen,  Pramana, 49, 609 (1997)
\bibitem{34}  S.  Xulu, S., Int. J. Theor. Phys., 39, 1153 (2000)
\bibitem{35}  V. Farhoni \& F.I. Cooperstock, F.I.,  Astrophys. J. 587, 483 (2003)




%









\end{thebibliography}
\end{document}